\RequirePackage{etex}

\documentclass[twocolumn]{aa}
\usepackage[varg]{txfonts}
\usepackage{natbib}
\usepackage{graphicx}
\usepackage{enumitem}
\usepackage{xcolor}
\usepackage{setspace}
\usepackage{underscore}
\usepackage{longtable}
\usepackage{amsmath}
\usepackage{amssymb}
\usepackage{nameref}
\usepackage{hyperref}
\usepackage{appendix}
\usepackage{makecell}
\usepackage{tabularx}
\usepackage{booktabs}
\usepackage{overpic}
\usepackage{multirow}
\usepackage{tikz}
\usepackage{textcomp}

\begin{document}

\title{Experimental evidence for a physical alteration \\of fine regolith on near-Sun asteroids}

\titlerunning{Physical alteration of regolith on near-Sun asteroids}
\author{Jooyeon Geem\inst{1}\thanks{Current address: National Astronomical Observatory of Japan, Mitaka 181-8588, Japan}
\and Mikael Granvik \inst{1,2}
\and Georgios Tsirvoulis \inst{1}
\and Leonard Schirner \inst{1}
\and Marianna L. Kemell \inst{3}
\and Antti Penttilä \inst{2}
\and Axel Hagermann \inst{1}
\and Athanasia Toliou \inst{1}
}
\institute{Asteroid Engineering Laboratory, Luleå University of Technology, Box 848, SE-981 28, Kiruna, Sweden
\and  Department of Physics, P.O. Box 64, 00014 University of Helsinki, Finland
\and Department of Chemistry, P.O. Box 55, 00014 University of Helsinki, Finland}

\date{Received 3 July 2026 / Accepted 27 October 2026}

    \abstract
  {The surfaces of airless bodies are directly exposed to the space environment and evolve through various processes such as (micro-)meteoroid impacts and thermal fatigue. These processes can modify physical properties such as the grain size, porosity, and mineral texture. Studying the physical properties can therefore provide clues as to the evolution of airless surfaces, and understanding them is also necessary for interpreting telescopic observations of the bodies. Near-Sun asteroids (NSAs), which approach the Sun at heliocentric distances of $\lesssim$0.2\,au, are notable targets in this regard, as their surfaces are likely modified by intense solar radiation. However, experimental studies of how radiative heating modifies the physical properties of carbonaceous asteroid surfaces remain scarce compared with studies of thermally induced changes in mineral composition and the resulting spectral alterations.}
  {This study experimentally investigates how intense radiative heating at near-Sun distances modifies the physical properties of carbonaceous asteroid surfaces.}
  {Near-Sun conditions at heliocentric distances of $0.09<r<0.19\,$au were simulated in the laboratory using the Space and High-Irradiance Near-Sun Simulator. CM-type carbonaceous chondrite simulants were irradiated under vacuum. Brightness changes were monitored in situ with a video camera during heating, and morphological changes were examined using a stage microscope and a scanning electron microscope (SEM).}
  {The irradiated samples show initial darkening followed by brightening, consolidation, and morphological modifications from sub-micrometer to millimeter scales. The observed brightness modifications are more consistent with changes in the absorption properties of the sample rather than variations in surface roughness. SEM imaging of the heated samples shows structures similar to those formed by sintering, which are completely absent in unheated samples.}
  {These results suggest that sintering is a likely candidate for the mechanism responsible for the observed consolidation and morphological changes. This work provides experimental evidence to help constrain the surface evolution of NSAs in general, and supports the sintering hypothesis as a possible mechanism proposed for (3200) Phaethon in particular.}


\maketitle
\nolinenumbers
\section{Introduction}

Small Solar System bodies are covered by regolith layers that vary in their physical properties. Here, physical properties refer to the structural characteristics of the regolith, such as its grain size, porosity, and surface texture rather than its chemical composition \citep{Cambioni+2021}. These physical properties provide clues about the processes that have occurred on these bodies. For example, ejecta-producing mechanisms such as outgassing and meteoroid impacts preferentially remove fine particles, leading to surfaces covered by coarser grains \citep{Cintala+1978, Gundlach+2013, Ballouz+2020}. Conversely, thermal fatigue driven by diurnal temperature variations can break down particles into smaller grains, producing fine particles \citep{Molaro+2017, Delbo+2014, Molaro+2020, Lucchetti+2024}. Furthermore, seismic activity transports fine particles from the surface into deeper regolith layers through size-sorting processes such as the Brazil nut effect \citep{Viranga+2016}. Thus, the physical properties of the regolith record the evolutionary history of the bodies \citep{Molaro+2020, Cambioni+2021}.

Knowledge of regolith physical properties is essential for interpreting telescopic observations of airless bodies. To date, small bodies have been extensively observed in the visible to near-infrared (VIS--NIR) wavelength range \citep{Bus+2002, Demeo+2009, Perna+2016, Binzel+2019, Popescu+2019, Lupishko+2018}, where the detected signal is dominated by sunlight scattered by the regolith layer. Such light scattering is affected by the physical properties and geometric albedo of the asteroid surface. Geometric albedo is an optical property of the surface that depends on both the mineral composition and physical structure. Consequently, telescopic observations (e.g., asteroid reflectance spectroscopy and polarimetry) are influenced not only by the mineral composition but also by the physical properties and albedo of the regolith \citep{Geake+1986, Geake+1990, Cloutis+2012}. Indeed, the grain size, porosity, packing density, and viewing geometry alone change the spectral slope and reflectance of carbonaceous chondrites in the laboratory \citep{Johnson+1973, Binzel+2015, Cloutis+2018, Potin+2019}. To properly interpret observational measurements, it is necessary to consider the physical properties of the regolith.

Near-Sun asteroids (NSAs) are noteworthy targets for studying regolith physical properties and their evolution. NSAs are asteroids with small perihelion distances ($q \lesssim 0.2$\,au; \citealt{Granvik+2016}), allowing them to approach the Sun closely and experience intense solar radiation heating. Such extreme conditions may significantly modify the physical properties of their surface regolith, potentially explaining several unresolved observational features.

For example, the largest known carbonaceous NSA, (3200) Phaethon, has a perihelion distance of $q = 0.14\,$au, and its surface temperature exceeds 1000~K at its perihelion \citep{MacLennan+2022}. Phaethon exhibits several unresolved features related to its physical properties. Its surface is expected to be dominated by coarse regolith grains (approximately hundreds of micrometers; \citealt{Ito+2018}), whereas the ejected dust consists of much finer particles ($\sim$1 {\textmu}m; \citealt{Jewitt+2013, Hui+2017}). In contrast, the Geminid meteoroid stream associated with Phaethon is composed of larger particles (approximately a millimeter; \citealt{Madiedo+2013, Blaauw+2017}). The mechanisms responsible for producing such seemingly inconsistent particle-size distributions remain unclear. In addition, Phaethon shows a blue visual spectral slope, for which the relative contributions of its mineral composition, regolith physical
properties, and inheritance from its parent family remain uncertain \citep{Yang+2010, deLeon+2010, Cloutis+2012, Lisse+2022}. One possible explanation is that intense solar radiation modifies the surface regolith. For example, \citet{Ito+2018} suggest that the presence of coarse particles may be related to sintering processes. In this context, it is necessary to study how intense solar radiation heating affects the physical properties of carbonaceous NSA regolith.

The effects of intense radiative heating on the physical properties of carbonaceous NSAs have not yet been systematically investigated in a laboratory setting. 
Although numerous experiments have examined the thermal processing of carbonaceous materials, most have focused primarily on compositional and spectral changes \citep[e.g.,][]{Ikramuddin+1975, Matza+1977, Akai+1992, Hiroi+1993, Hiroi+1996, Nakato+2008, Mogi+2017, Matsuoka+2022, Sidhu+2023, Yu+2024, Zheng+2024}. In comparison, relatively few studies have examined changes in the physical properties, mainly focusing on grain-size modification driven by thermal stress during temperature cycling \citep[e.g.,][]{Delbo+2014, Libourel+2020, Patzek+2024, Latsia+2026}. Despite these contributions, several key aspects remain insufficiently explored. To our knowledge, all of the above experiments relied on furnaces, conductive heating stages, laser irradiation, or climatic chambers rather than solar-like radiative heating. In addition, among the studies that examined the physical properties \citep[i.e.,][]{Delbo+2014, Libourel+2020, Patzek+2024, Latsia+2026}, all but the one by Patzek et al.\ were conducted under ambient pressure rather than under vacuum.

As a result, there remains a knowledge gap in understanding how intense solar-like radiative heating under vacuum affects the physical evolution of carbonaceous asteroidal materials, limiting a direct comparison between laboratory results and telescopic observations of carbonaceous NSAs such as Phaethon. For example, \citet{Ito+2018} propose that the coarse surface grains on Phaethon may form through sintering, but there are no experimental constraints on whether sintering can occur under Phaethon’s surface conditions.

In this study, we investigate with laboratory experiments how intense radiative heating in the near-Sun environment affects the surface properties of carbonaceous asteroid regolith. We focus on the physical properties of the regolith, such as surface texture and grain-size evolution. We also examine its brightness, which can be modified by changes in both its physical properties and its mineral composition. We describe our experimental setup, sample preparation, and image-analysis techniques in Sect. \ref{sec:method}, and the observed results in Sect. \ref{sec:result}. We then discuss the limitations of our experiments and the possible mechanisms responsible for the observed phenomena in Sect. \ref{sec:discussion}. Finally, we offer our conclusions in Sect.~\ref{sec:conclusions}.

\section{Methods}
\label{sec:method}
\subsection{Instrument}
\label{sec:method-instrument}
To simulate the vacuum and the electromagnetic radiation in the near-Sun environment, we employed the Space and High-Irradiance Near-Sun Simulator (SHINeS; see \citealt{Tsirvoulis+2022} for a detailed description of the instrument). We used a Xenon-arc lamp as the radiation source, a common choice for solar simulators. We adjusted the irradiation intensity by varying the height of the sample stage and tuning the lamp power. This combination allowed us to simulate radiative intensities equivalent to heliocentric distances $0.09<r<0.19\,$au. The light spot is 35~mm in diameter at the sample plane \citep{Tsirvoulis+2022}, so the entire sample (20~mm in diameter) is illuminated. The irradiance is nearly uniform over the brightest central region of the spot, which \citet{Tsirvoulis+2022} define as a circular area about
13~mm in diameter, and decreases gradually toward the edge. The top of the vacuum chamber has a viewport through which a high-speed video camera (Chronos 1.4, KRON Technologies), operated in monochromatic mode, monitored the sample during heating at 60 frames per second (FPS), allowing us to track variations in the sample, such as brightening. Based on the spectral distribution of the incident light \citep{Tsirvoulis+2022} and the spectral response of the camera in monochrome mode\footnote{\url{https://www.krontech.ca/wp-content/uploads/2021/09/FM-ENGR-50001-Chronos-1.4-Datasheet-Rev5.pdf}}, the observed signal is expected to be broadly representative of the Johnson-Cousins $V$ band. We did not use astronomical filters.

\subsection{Sample preparation} \label{sec-sample}
Our sample is the CM carbonaceous chondrite simulant developed by the Exolith Lab at the University of Central Florida \citep{Britt+2019}. We sieved the fresh simulant powder using a 45~{\textmu}m mesh. According to \citet{Britt+2019}, the grain sizes of the raw materials used in the simulant range from microns to tens of microns. Because minerals with larger grain sizes could be preferentially retained on the sieve, sieving may introduce a compositional bias. To test this, we prepared an additional sample in which the retained fraction was ground with a mortar and pestle until it passed entirely through the mesh and was then recombined with the sieved powder. The experimental results with these fully ground samples were identical to those obtained with the original simulant powder (i.e., without additional grinding). Therefore, we believe that any mineral bias introduced by the 45~{\textmu}m sieving process is negligible. This work mainly focuses on fine particles ($<45\,\mu\mathrm{m}$), except that we used larger particles for the additional experiment (Sect. \ref{sec:consolidation}).

The sieved samples were then placed in an aluminum-oxide holder with an inner size of 2 × 2 cm and a height of 0.5 cm. We then lightly tamped the sample plate several times to flatten the sample surface. In this way, the sample surface generally became flat, while cohesive forces between the fine particles created small lumps, resulting in surface roughness on the millimeter scale (Fig.~\ref{fig:morpholoty_cm2um}a).

\subsection{Heating process}
\label{sec:Method-heating process}

The heating treatment process consisted of the following steps. First, to measure the light intensity, we turned on the lamp and waited for tens of minutes until the lamp reached a stable temperature. During this time, the shutter remained closed, so the light did not reach the sample stage inside the chamber. After preheating, a light sensor (PRONTO-500 from Gentec Electro-Optics) was placed on the sample stage at the exact position where the sample would be located. The shutter was then opened to measure the light intensity to which the sample would be exposed. This measurement was repeated at least three times to ensure consistency in measurement. Second, the sample was placed on the stage inside the chamber. The chamber was then pumped down until the pressure reached $\sim10^{-5}$~mbar. The sample generally rested in the vacuum chamber for at least 4 hours prior to heating. After the lamp was turned on and reached a stable temperature, heating began by opening the shutter. If the heating time exceeded two hours, the process was paused for several tens of minutes to prevent overheating of the equipment, and heating was resumed. During the whole irradiation process, the sample was recorded by the high-speed camera. Finally, after the heating was complete, the sample was left in the vacuum chamber for one day to cool down. This cooling period under vacuum prevented the sample from being oxidized by ambient air while the sample was still at a high temperature after heating. Then, the sample was taken out of the chamber and analyzed by, for example, a microscope.    

\subsection{Microscopy}
We examined the physical structure of the samples across multiple size scales using both a stage microscope and a scanning electron microscope (SEM). The stage microscope was used to observe particle structures from the millimeter scale down to several hundred micrometers. Images were acquired using an AmScope stereo microscope. The pixel scale at each magnification was calibrated using a stage micrometer (R1L3S2P by Thorlabs). The spatial resolution of the stage microscope was approximately tens of micrometers. Sample images were acquired both before and after heating. The microscope was placed a few meters from the vacuum chamber to minimize any changes during sample transport.

An SEM was used to analyze the structures from the sub-micrometer scale up to several hundred micrometers. We acquired SEM images using a field-emission scanning electron microscope (S-4800 by Hitachi High-Technologies) equipped with an Oxford INCA 350 energy-dispersive X-ray spectrometer (EDS) at the University of Helsinki. The samples were mounted on aluminum stubs using conductive carbon tape and sputter-coated with a thin Au/Pd layer prior to imaging to prevent charging. Secondary electron images were acquired at 5.0~kV, and EDS measurements were conducted at 20~kV. The samples were transported from Kiruna to Helsinki for SEM analysis. Although fragmentation into millimeter-sized pieces occurred during transport, the sub-micrometer-scale microstructure was expected to be preserved because the fragmentation did not involve substantial deformation or pulverization.

\section{Results}
\label{sec:result}
By exposing the CM simulants to intense solar-like radiation, we observed several effects including brightness changes, consolidation, and morphological changes from the sub-micrometer to the millimeter scale. The following subsections describe each phenomenon in detail.

\subsection{Brightness variations}
\begin{figure}
    \centering
    \includegraphics[width=0.95\linewidth]{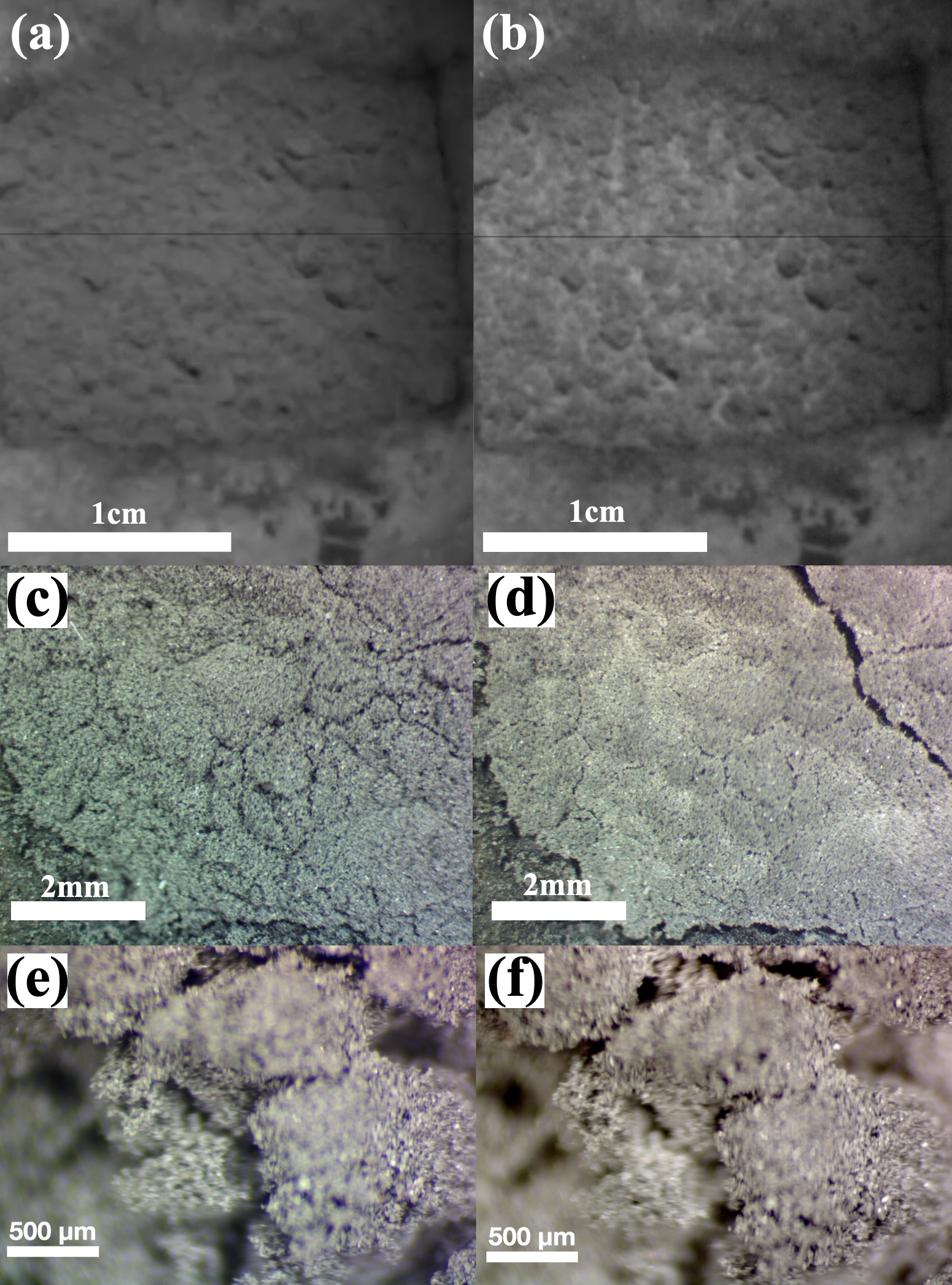}
    \caption{Comparison of samples before and after heating. For all panels, the left column shows the samples before heating, and the right column shows them after heating. The equivalent heliocentric distance of the experiments is  $r\sim0.12\,$au. Panels (a) and (b) are high-speed camera images with identical contrast ranges. Panels (c)–(f) are stage microscope images at different magnifications. Scale bars are shown in the lower-left corner of each image. Colors and brightness in panels (c)–(f) have been adjusted for visibility. The entire sample is illuminated by the light spot. See Sect.~\ref{sec:method-instrument} for the size and uniformity of the illuminated area.}
    \label{fig:morpholoty_cm2um}
\end{figure}

Brightening of the CM simulant samples occurs upon radiation heating under vacuum conditions, and becomes more pronounced with increasing radiative heating (i.e., decreasing heliocentric distance $r$). Such brightening indicates an increase in the surface reflectance of the samples. 
We observed brightening after radiative heating at $0.09<r<0.19\,$au, although for samples heated at $r \gtrsim 0.18\,$au the brightening was barely detectable. Figures~\ref{fig:morpholoty_cm2um}a and b show the sample before and after heating. At first glance, the entire sample appears brighter after heating, but a closer look reveals that the brightening is spatially heterogeneous.

On the unheated sample surface, fine particles aggregate through cohesive forces, forming a rough structure with valleys and peaks. Before heating, these valleys and peaks showed similar brightness. However, after heating, the valleys became brighter than the peaks. To isolate the effect of surface roughness, we conducted a separate experiment. We removed the initial roughness by pressing the sample surface flat, and then made artificial pits. After heating, brightening occurred uniformly on the flat areas, whereas the interiors of the pits brightened more strongly. This demonstrates that the initial surface roughness can affect the degree of brightening. For example, in the valley regions, multiple scattering traps incident light and enhances its absorption, increasing local heating and thus brightening.

\begin{figure}
\centering
\includegraphics[width=\linewidth]{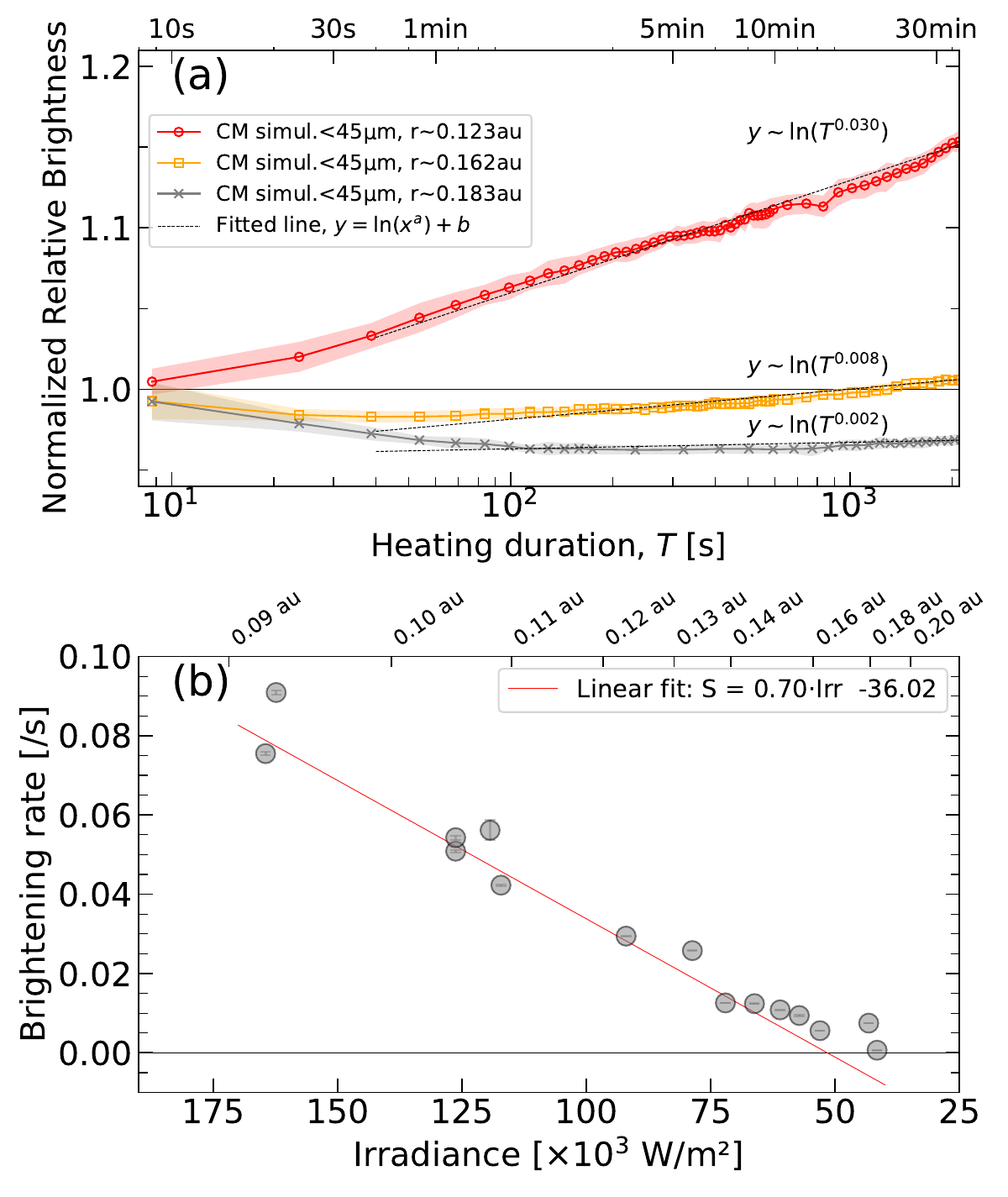}
\caption{Brightness evolution of the CM simulant ($<45$~{\textmu}m) under radiative heating. (a) Normalized relative brightness evolution over the heating duration for samples with grain size smaller than 45~{\textmu}m. The dashed line is fitted to the linear part of the brightness evolution. (b) Relationship between the brightening rate ($a$) and the corresponding heliocentric distance ($r$).}
\label{Fig:brightening}
\end{figure}

To quantify the evolution of brightening, we measured the temporal change in sample brightness during heating (Fig.~\ref{Fig:brightening}a). The geometry was fixed during each experiment, with the radiation incident nearly normal to the sample surface ($i \sim 0^\circ$) and the camera at an emission angle of $e \sim 30^\circ$ (phase angle $\alpha \sim 30^\circ$). Between experiments, $e$ varied by at most $4^\circ$ due to differences in the height of the sample table. Thus, the observed brightness changes are unlikely to result from changes in geometry.

In this paper, ``brightness'' refers to all quantities derived only from the camera images, namely the pixel values and their normalized values. Brightness was determined in two normalization steps. First, the pixel values averaged within selected sample areas in each video frame were divided by the averaged pixel values of the sample holder in the same frame. Because the sample and the holder are illuminated by the same lamp and recorded in the same frame, this ratio removes temporal variations in the lamp intensity, assuming that the reflectance of the holder does not change during heating. Second, this ratio was divided by its average over the first 10~s of heating. Because the geometry was fixed during each experiment, this second step cancels the remaining constant factors, such as the reflectance of the holder and the viewing geometry. The resulting normalized brightness therefore traces the bidirectional reflectance \citep{Hapke+2012} of the sample over time, relative to its value at the onset of heating, and can be compared directly between experiments. The standard deviation of the normalized brightness within the sample areas is shown as a shaded region in Fig.~\ref{Fig:brightening}a.

Representative examples of brightness evolution are shown in Fig.~\ref{Fig:brightening}a for three levels of radiative heating ($r = 0.12$, $0.16$, and $0.18\,$au). For small $r$, the brightness increases immediately, and continues to increase logarithmically with time. At intermediate $r$, an initial darkening occurs during the first few minutes, followed by brightening. At the largest $r$, only the initial darkening is observed, with little or no subsequent brightening. The final degree of brightening decreases with increasing $r$.

This behavior may reflect a temperature-dependent transition. Because the sample heats up gradually rather than reaching the target temperature instantly, it passes through a low-temperature stage early in each run. We infer that darkening dominates below a certain temperature threshold and brightening above it. At the largest $r$, the sample may never cross the threshold, leaving only the initial darkening. At the smallest $r$, the temperature rises rapidly, making the darkening phase too short to detect.

To investigate the dependence of brightening on heliocentric distance, we fitted the brightness evolution using $y = a \ln(x) + b$, where $y$ is the normalized brightness and $x$ is the heating duration. For experiments exhibiting initial darkening, only the brightening phase was used for fitting. In Fig.~\ref{Fig:brightening}a, the fitted line is shown by the dashed line. The fitted parameter $a$ was adopted as a measure of the brightening rate. Applying this procedure to experiments conducted under different radiative fluxes, we found that the brightening rate $a$ shows a linear correlation with increasing flux over the range corresponding to $0.09<r<0.19\,$au (Fig.~\ref{Fig:brightening}b). The red solid line shows a linear regression of the brightening rate as a function of radiative flux, performed on binned data to mitigate biases from uneven sampling.

\begin{figure*}[h]
  \centering
  \begin{overpic}[width=0.48\textwidth]{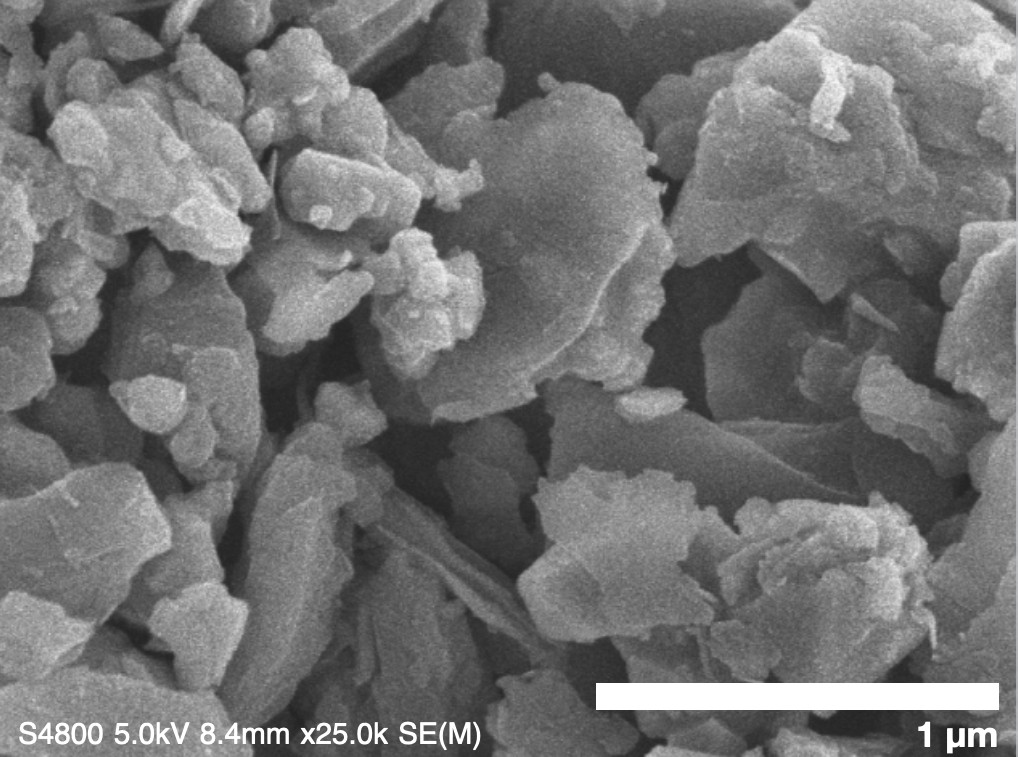}
    \put(3,67){\Large\textcolor{white}{\textbf{(a)}}}
     \put(50,70){\makebox(0,0){\large\textcolor{white}{\textbf{Before heating}}}}
  \end{overpic}
  \hfill
  \begin{overpic}[width=0.48\textwidth]{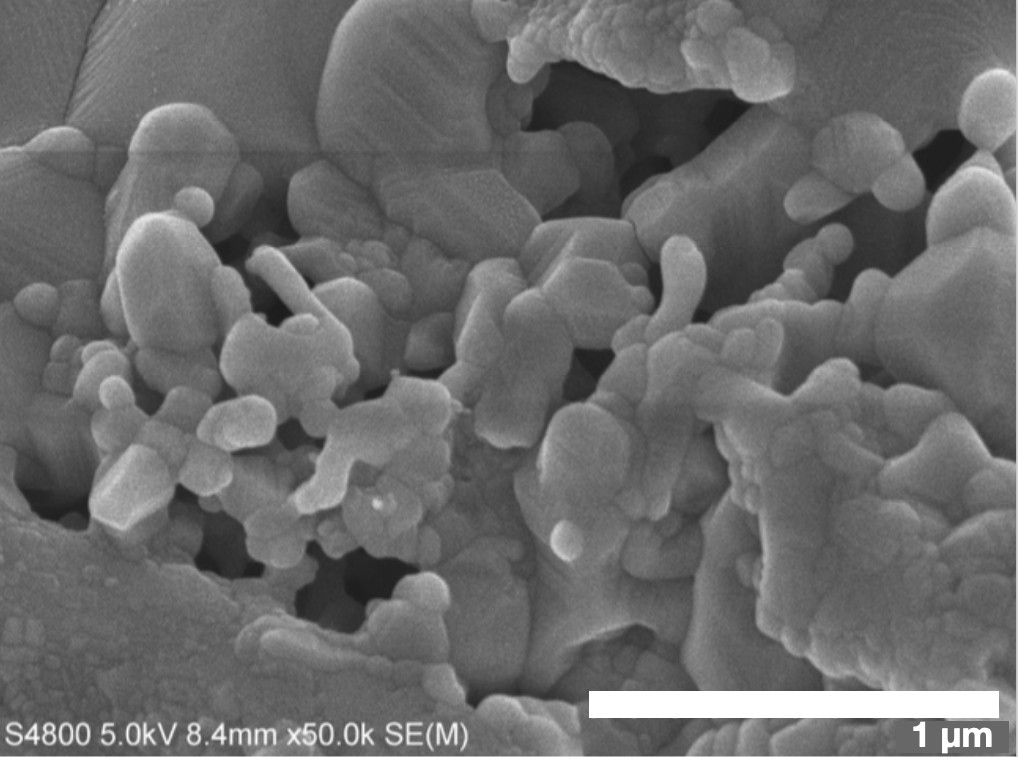}
\put(3,67){\Large\textcolor{white}{\textbf{(b)}}}
\put(50,70){\makebox(0,0){\large\textcolor{white}{\textbf{After heating}}}}
\end{overpic}
  \vspace{0.5cm}

  \caption{SEM images of the CM simulant sample (sieved to <45~{\textmu}m) before and after heating at $r = 0.11\,$au. Figure (a) shows the sample prior to heating,
whereas (b) shows the sample after 1.5~h of heating. Scale bars are shown in the lower-right corner of each image. Note that the images of the samples before and after heating were not taken from the same location. The two images have been rescaled so that the scale bars are identical in length, allowing direct comparison.
}\label{fig:SEM}
\end{figure*}

\subsection{Morphology}

Morphological changes in the heated sample were observed across a wide range of spatial scales ranging from sub-micrometers to millimeters. At the millimeter scale, the surface became noticeably smoother after heating. Figures~\ref{fig:morpholoty_cm2um}c and d show stage-microscope images of the same area before and after heating. Before heating, fine particles aggregated through cohesive forces, forming clumps and creating a rough surface with peaks and valleys. After heating, some of these valleys were filled in, resulting in a flatter surface. The sample also appeared to shrink, as indicated by gaps that formed between the sample and the sample holder, as well as by a crack visible in the upper-right region of Fig.~\ref{fig:morpholoty_cm2um}d. At the scale of several hundred micrometers, we could not identify distinct structural differences resolved by our stage microscope (approximately tens of micrometers), as shown in Figs.~\ref{fig:morpholoty_cm2um}e and f.

At the sub-micrometer scale, heating produced distinct morphological changes (Fig.~\ref{fig:SEM}). Before heating, the mineral boundaries at the sub-micrometer scale were sharp, and the minerals exhibited distinctly angular textures. Mg-serpentine, which accounts for $\sim 70$\,wt\% of the sample \citep{Britt+2019}, appeared in a platy morphology, showing the typical layered phyllosilicate structure resembling a stacked pastry. Olivine ($\sim$11\,wt\%) and magnetite ($\sim$5\,wt\%) occurred as angular aggregates.  After heating, however, these angular boundaries mostly disappeared, and the overall surface became rounder and smoother. In Fig.~\ref{fig:SEM}b, the tips of the serpentine grains appear curled and less angular. In all SEM images of the heated sample, we consistently observed particles that appear to stick together, with necks formed between them. In particular, structures resembling particle bonding were most commonly observed among small particles (i.e., magnetite and olivine).

\begin{figure}
    \centering
    \includegraphics[width=\linewidth]{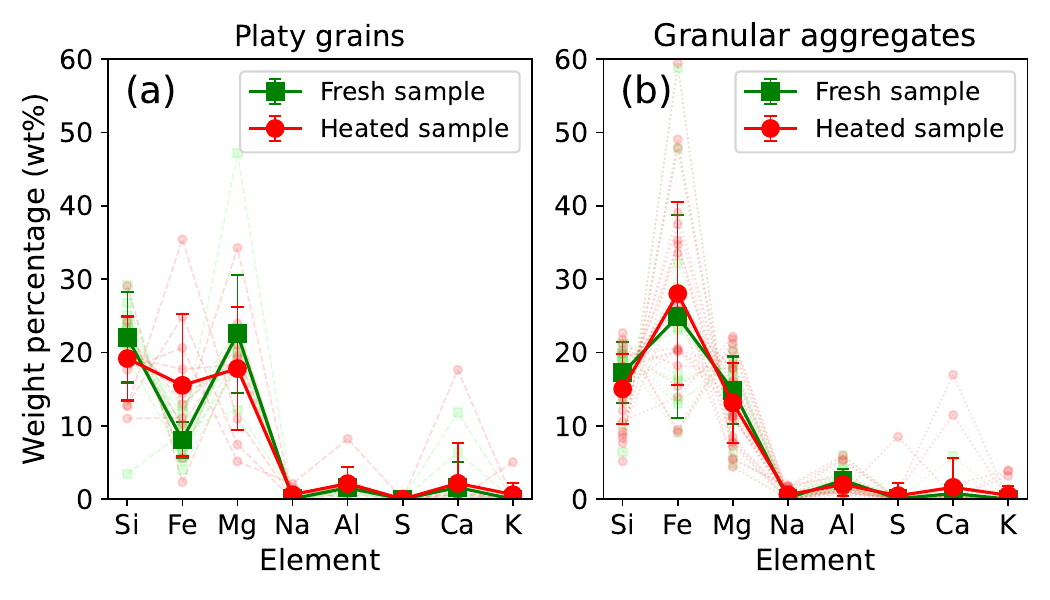}
    \caption{Energy-dispersive X-ray spectroscopy (EDS) analyses of both fresh and heated samples. Dashed lines indicate measurements from different locations within the samples, and the solid lines indicate the average values from those locations. Green square and red circle markers represent the fresh and heated samples, respectively. (a) Measurements from platy particles likely corresponding to phyllosilicates (or dehydrated phyllosilicates in the heated sample). (b) Measurements from granular particles likely corresponding to olivine or magnetite.}
    \label{fig:EDS}
\end{figure}

One possible concern is that the SEM images do not compare identical locations before and after heating, thus the observed morphological changes could reflect mineralogical heterogeneity rather than heating effects. To address this potential sample bias, we conducted EDS analyses of unheated and heated samples. Platy and granular particles were analyzed separately using tens of measurements for each type. The averaged relative elemental compositions of the samples are shown in Fig.~\ref{fig:EDS}. 

Based on the EDS results, we conclude that the observed morphological changes are caused by heating rather than sample bias. In Fig.~\ref{fig:EDS}, the granular particles show no significant change in relative elemental composition between the unheated and heated samples, indicating that the same mineral population is being compared. Considering their morphology, composition, and the simulant recipe of \citet{Britt+2019}, we interpret these granular particles as a mixture of magnetite and olivine. Because both minerals are refractory, compositional changes are not expected under our experimental conditions, consistent with the EDS observations. The platy particles also show broadly similar elemental compositions before and after heating, except for an increase in the relative Fe abundance in the heated sample. These platy particles are likely Mg-serpentine, which undergoes dehydration at temperatures above several hundred degrees Celsius \citep{Hiroi+1993}. Thus, the apparent increase in the relative Fe abundance may be associated with the loss of volatiles during this serpentine dehydration process.

\subsection{Consolidation}
\label{sec:consolidation}

\begin{figure}
  \centering

  \begin{overpic}[width=0.65\columnwidth]{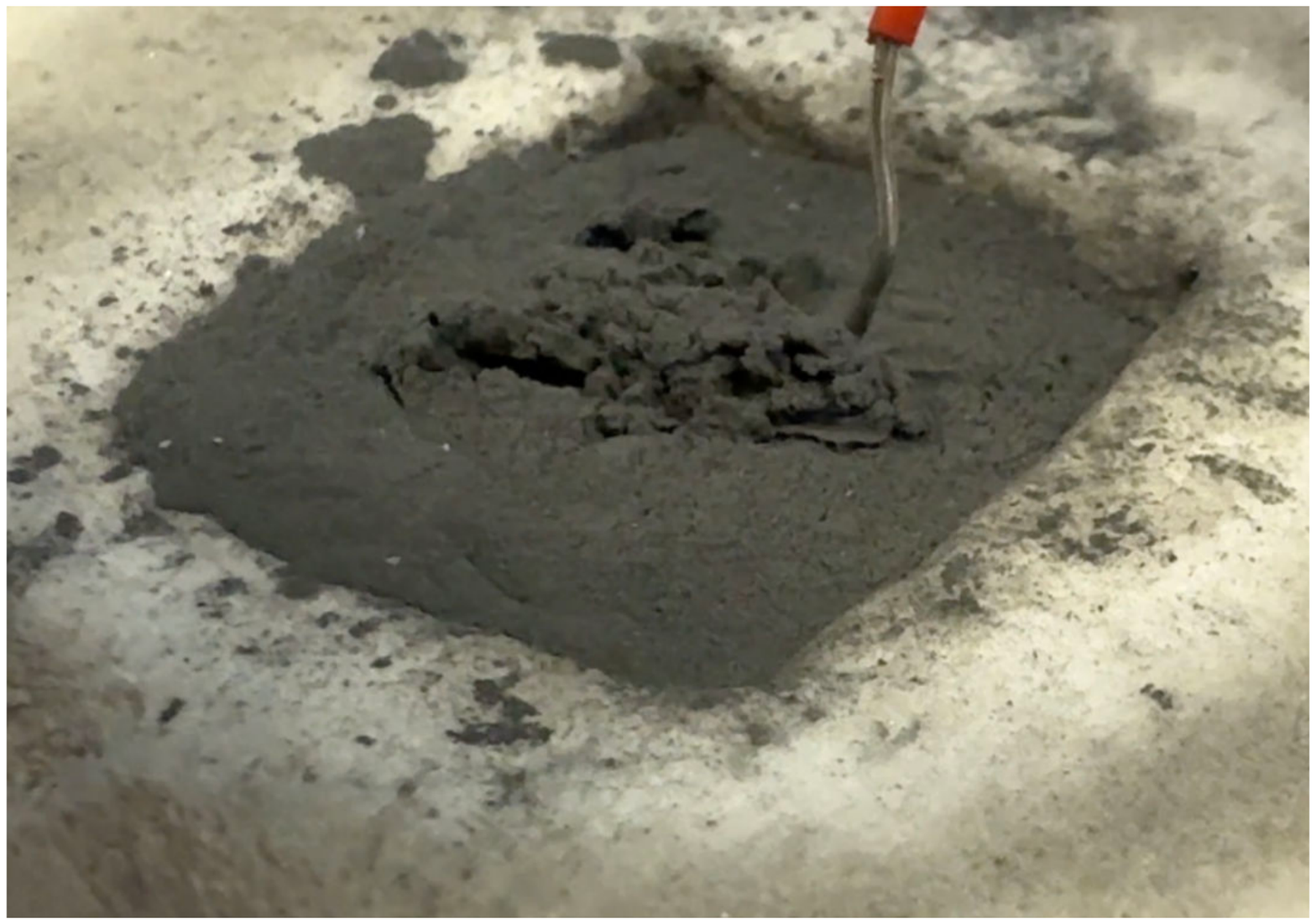}
    \put(6,60){\large\textbf{(a)}}
  \end{overpic}
  \vspace{0.5em}
  \begin{overpic}[width=0.65\columnwidth]{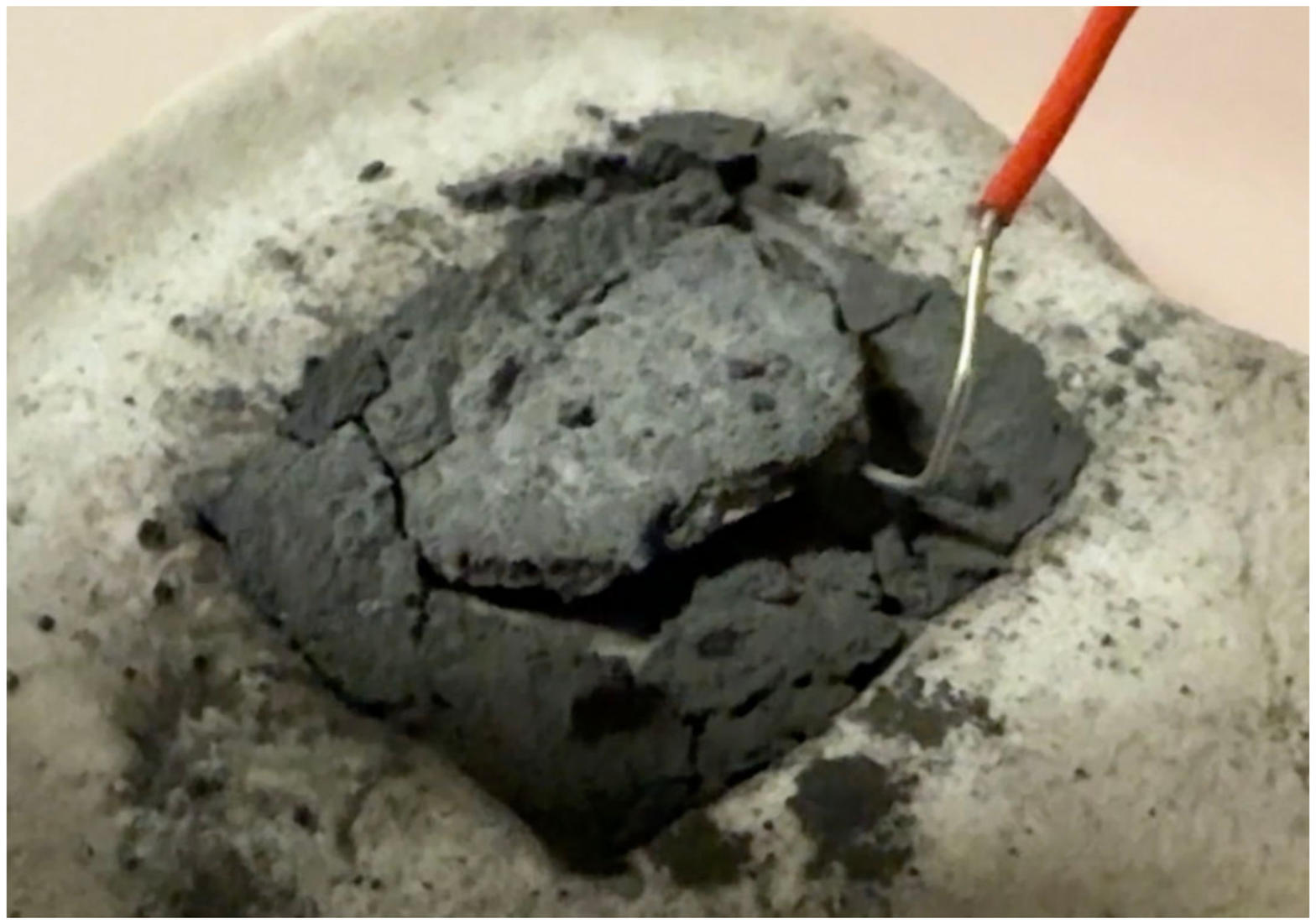}
    \put(6,60){\large\textbf{(b)}}
  \end{overpic}
  \caption{Consolidation of the CM simulant ($<45$~{\textmu}m) by radiative heating. (a) Example image of the sample before heating. The sample cannot be lifted using a wire. Instead, sample particles agglomerate and form clumps due to cohesive forces.
  (b) The same sample as in image (a), but after heating. 
  Consolidation occurs to transform the sample into a plate-like structure. However, the consolidated structure remains fragile and is easily fractured. The associated movies are available online.}
  \label{fig:Consolidation}
\end{figure}

By visual inspection, we found that heating induced consolidation in the fine-grained ($\lesssim 45$~{\textmu}m) CM simulant. In the unheated sample, the fine particles stuck together to form clumps due to cohesive forces, but the sample was not sufficiently consolidated. When we attempted to lift it with a wire, it crumbled into small clumps and loose powder (Fig.~\ref{fig:Consolidation}a).

In contrast, the heated samples were sufficiently consolidated to be lifted as a single intact piece (Fig.~\ref{fig:Consolidation}b). This indicates that consolidation occurred due to the heating process. While these qualitative observations (i.e., based on whether a sample could be lifted intact with a wire) suggest an increase in mechanical strength after heating, quantitative measurement was not possible due to the limitations of available mechanical testing methods in this study. Systematic measurement of the strength is left for future work. Movies showing this phenomenon are available online.

In addition, the presence of fine particles appears to play a key role. We found that consolidation observable by the naked eye occurred only when fine particles were present in the sample. We conducted additional experiments using different grain-size fractions, including coarse particles (1000~{\textmu}m sieve), medium particles (710~{\textmu}m sieve), fine particles (45~{\textmu}m sieve, i.e., the primary sample used in this study), and their mixtures. Samples with higher fractions of fine particles showed stronger consolidation and could be lifted intact. In contrast, samples containing larger fractions of coarse particles tended to crumble when lifted or could not be lifted at all. Samples composed only of fine particles remained intact when lifted (Fig.~\ref{fig:Consolidation}b), whereas samples consisting only of coarse particles showed no visible consolidation. Nevertheless, we emphasize that the consolidated sample was brittle, with a strength comparable to that of a dried mud crust. It retained its shape when lifted but snapped easily when touched.

\section {Discussion}
\label{sec:discussion}

This study simulates a near-Sun environment in the laboratory to investigate how the surfaces of NSAs evolve when they are exposed to strong solar electromagnetic radiation. Using CM simulants, we examined changes in their physical properties and brightness induced by intense radiation heating. As a result, we observed several physical and morphological changes, including their brightening, darkening, consolidation, and morphological modifications spanning from the millimeter scale down to the sub-micrometer scale after the heating process.

Before further discussing our experimental results, we note several limitations of this study. 
First, microgravity conditions were not reproduced in our experiments. However, we expect the behavior of the sample particles in our laboratory to be similar to that under microgravity because of the small particle size ($<45$~{\textmu}m) used in our experiments. \citet{Scheeres+2010} showed that for sufficiently small particles, cohesive forces dominate over gravity, making terrestrial experiments on cohesive powders relevant for understanding processes on asteroid surfaces. On Earth, the particle size at which gravitational weight equals cohesive force is $\sim 650$~{\textmu}m \citep{Scheeres+2010}. The particles used in our experiments ($<45$~{\textmu}m) are below this threshold, supporting the relevance of our results to surface processes on NSAs. 
Second, the composition of our simulants does not exactly match that of real asteroids. For example, coal was used as a substitute for meteoritic insoluble organic matter \citep{Britt+2019}. However, as discussed below in Sect.~\ref{sec:dis-brightening}, results from experiments with CM meteorites are consistent with those from our simulant samples under the same experimental conditions, which supports the relevance of our work employing simulants. Third, our work was conducted using CM simulants and a CM meteorite (Aguas Zarcas). Thus, it remains unclear if the observed behaviors are applicable to other primitive materials (e.g., other carbonaceous chondrites) or depend on specific compositions, leaving this for future work. Fourth, the heating rate in our experiments may be faster than that experienced by actual asteroids. In our setup, samples were heated to several hundred degrees Celsius within a few minutes after the irradiation began. This timescale is likely shorter than that in real asteroidal environments. For example, surface temperatures on Phaethon are estimated to increase by only several tens of degrees over a few minutes \citep{MacLennan+2024}. This study also does not account for diurnal temperature cycles. However, our focus is on the effects of peak temperature on physical changes in this study, and experiments that consider time-dependent thermal evolution should be investigated in future work. In addition, our analysis is based on particles smaller than 45~{\textmu}m. Larger particles were measured only in a supplementary experiment (Sect. \ref{sec:consolidation}). The influence of larger particle sizes remains to be explored.

\subsection{Mechanism of brightness variations}
\label{sec:dis-brightening}

In our experiments, samples exposed to strong radiation ($0.09<r<0.19\,$au) became brighter after heating, whereas brightening was negligible at $r\gtrsim 0.18\,$au. In addition, we observed that some samples showed initial darkening during the first few minutes of heating, especially at larger $r$ (i.e., lower radiative flux), followed by subsequent brightening. The final degree of brightening was found to correlate linearly with the radiation intensity.

The initial roughness of a sample (i.e., peaks and valleys) affects the degree of brightening and produces its spatial heterogeneity within the sample. However, we interpret the observed brightness change as being primarily driven by modifications in the absorption properties of the samples, rather than by changes in surface roughness at the sub-micrometer to micrometer scale. Several mechanisms for thermally induced darkening and brightening of carbonaceous materials have been proposed in previous studies, including carbonization, thermal decomposition of opaque carbonaceous components, dehydration of phyllosilicates, and modification of surface roughness \citep[e.g.,][]{Hiroi+1993, Hiroi+1996, Moroz+1998, Moroz+2004, Gillis-Davis+2015, Lantz+2015, Laczniak+2020, Sidhu+2023, Laczniak+2024}. Among these, we suggest that carbonization, thermal decomposition of opaque carbonaceous components, and dehydration of phyllosilicates are particularly relevant to our experimental results, as discussed below.

Carbonization and thermal decomposition of carbonaceous materials are closely associated with pyrolysis, defined as the thermal decomposition of organic materials under oxygen-free conditions \citep[][and references therein]{Devi+2021}. In the early stages of pyrolysis ($T\lesssim600\,^\circ\mathrm{C}$), relatively weak covalent bonds (e.g., C–H and C–O) are preferentially cleaved, releasing gases such as H$_2$O, CO, CO$_2$, and hydrocarbons \citep{Biron+2016}. The progressive loss of hydrogen- and oxygen-bearing functional groups results in carbon enrichment and carbonization of the samples. In addition to these chemical changes, structural changes begin once the temperature exceeds $\sim 600\,^\circ\mathrm{C}$. At such temperatures, the basic structural units (BSUs), initially small and randomly oriented polyaromatic layers, progressively align and stack into more ordered structures \citep{Oberlin+1984}. These structural changes are known to modify the electronic absorption properties and increase reflectance \citep{Moroz+1998, Moroz+2004, Rouzaud+1989}. Phyllosilicates also contribute to the reflectance change of our samples. \citet{Hiroi+1999} showed that lizardite darkens at $T\lesssim400$--$500\,^\circ\mathrm{C}$ and brightens at higher temperatures. Lizardite is one of the major phyllosilicates present in our CM simulants \citep{Britt+2019}.

Based on the mechanisms above, we interpret the darkening to brightening transition observed in our samples as follows. The initial darkening is likely caused by carbon enrichment resulting from thermal decomposition and devolatilization of carbonaceous materials, together with the darkening of phyllosilicates (i.e., lizardite in our simulant). As the sample temperature increases beyond $\sim600\,^\circ\mathrm{C}$, structural reorganization of carbonaceous materials begins, leading to an increase in reflectance (brightening). Furthermore, lizardite brightens at temperatures of $T \gtrsim 400\,^\circ$C \citep{Hiroi+1999}, which may also contribute to the observed brightening of our samples. Because our experiment does not measure the sample temperature in real time, we cannot determine the exact temperature at which this transition occurs. However, \citet{Hiroi+1993} reported that CM chondrites darken upon heating and then brighten once the heating temperature exceeds $\sim600\,^\circ\mathrm{C}$. This is consistent with our interpretation and suggests that our samples reach $\sim600\,^\circ\mathrm{C}$ around the darkening-to-brightening transition.

Our results are broadly consistent with previous heating experiments on carbonaceous materials. Like \citet{Hiroi+1993}, \citet{Moroz+2004} found that laser irradiation brightens complex hydrocarbons in the visible wavelengths. In contrast, \citet{Sidhu+2023}, who heated CM meteorites and their simulants over a relatively long period (several days), report the opposite temperature trend, in which samples became brighter up to $\sim$600--800\,$^\circ$C but darker at higher temperatures. This discrepancy may reflect differences in heating duration and method, and requires further investigation.

\begin{figure}
    \centering
    \includegraphics[width=\linewidth]{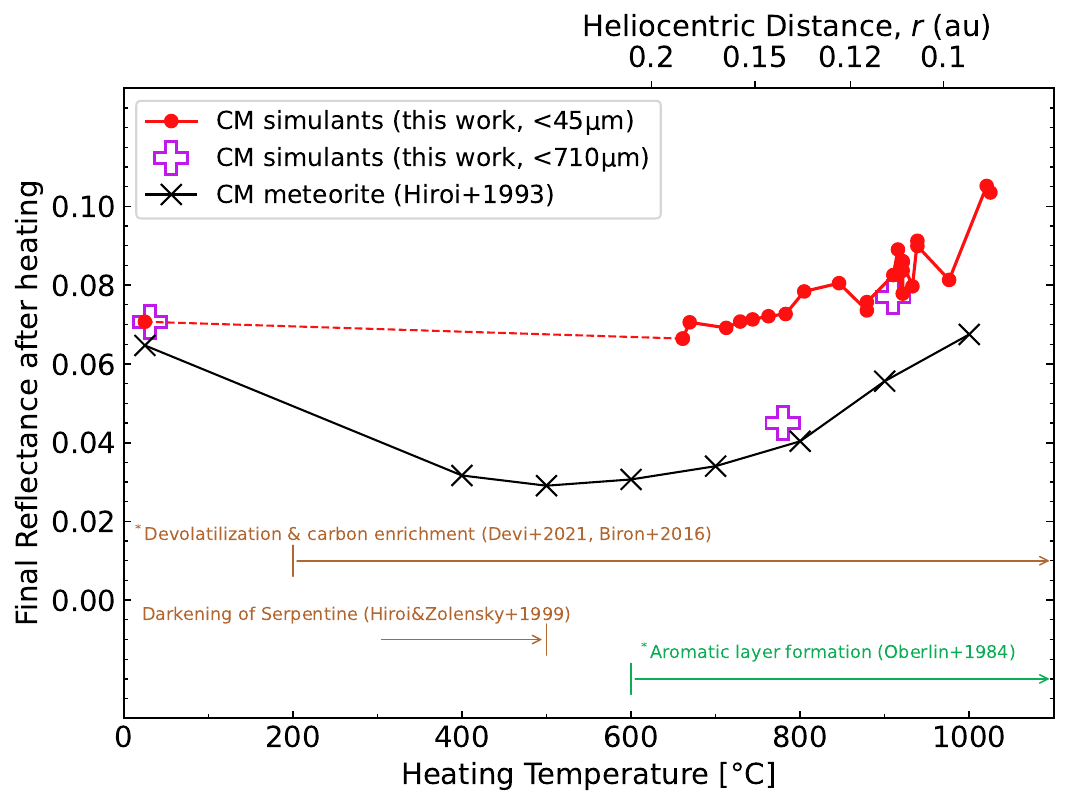}
    \caption{Final reflectance as a function of the heating temperature. 
    Red circles represent our CM simulants after heating, measured from the camera. Purple plus symbols indicate the reflectance of our simulants measured with the spectrometer (grain size sieved with a 710~{\textmu}m mesh), where reflectance is defined as the reflectance at 0.55~{\textmu}m. CM-type meteorites from \citet{Hiroi+1993} are also shown for comparison. The dashed line indicates the temperature range for which no measurements are available due to the limited range of $r$ achievable by SHINeS.}
    \label{fig:temp-brightness}
\end{figure}

Figure~\ref{fig:temp-brightness} compares our results with the CM-chondrite heating data of \citet{Hiroi+1993}. The horizontal axis shows the sample temperature (converted from $r$, see below), and the vertical axis is the final reflectance after heating. 
Each red circle is a single heating experiment. Because these are final values, the transient darkening observed early in each run is not shown here. Our main simulant ($<45$~{\textmu}m) covers only the high temperature regime ($T \gtrsim 600\,^\circ\mathrm{C}$) and so does not sample the darkening phase, but it shows a clear increase in reflectance with temperature, 
with a slope similar to that reported by \citet{Hiroi+1993}.

The two axes in Fig.~\ref{fig:temp-brightness} are derived as follows. Because our setup records the radiation intensity (i.e., the corresponding heliocentric distance $r$) rather than the sample temperature, we converted $r$ to an expected temperature using the surface-temperature model for near-Earth asteroids (NEAs) of \citet{Marchi+2009}, assuming a flat surface in instantaneous thermal equilibrium with sunlight and a geometric albedo of 0.05. The final reflectance was obtained by scaling the normalized brightness from the camera images by the reflectance of the unheated sample at 0.55~{\textmu}m. This reference value (0.076) was measured with the integrating-sphere UV–Vis–NIR spectrometer at the University of Helsinki \citep{Penttila+2018}. We also include measurements of a coarser simulant ($<710$~{\textmu}m, purple markers), whose reflectance was measured directly as absolute reflectance with the spectrometer rather than from camera images. The offset between the two simulant datasets likely reflects differences in grain size, measurement methods, and experimental conditions.

Modification of surface roughness is another mechanism proposed to explain thermally induced brightening (or darkening). However, we consider this mechanism unlikely to account for our experimental results. Previous studies \citep[e.g.,][]{Moroz+2004, Thompson+2019, Thompson+2020, Frattin+2025} have shown that surface roughness can influence the reflectance of a sample. For example, \citet{Frattin+2025} demonstrate that roughness variations on the order of tens to hundreds of micrometers affect the albedo of the Mojave Mars Simulant. In our samples, no detectable changes occurred at the tens- to hundreds-of-micrometers scale. In addition, after artificially modifying the surface roughness of the heated sample by pressing the surface or scratching it with a needle, the sample remained bright, indicating that the observed brightening is unlikely to result from changes in macroscopic surface roughness. Our interpretation is consistent with that of \citet{Moroz+2004}, who conducted ion irradiation experiments on natural complex hydrocarbons. The authors suggested that although ion irradiation may affect the surface micro-roughness, the thermally induced brightening observed in their samples results from changes in the absorption coefficients caused by modifications of the chemical structure of complex organics under irradiation.

\begin{figure}
    \centering
    \includegraphics[width=\linewidth]{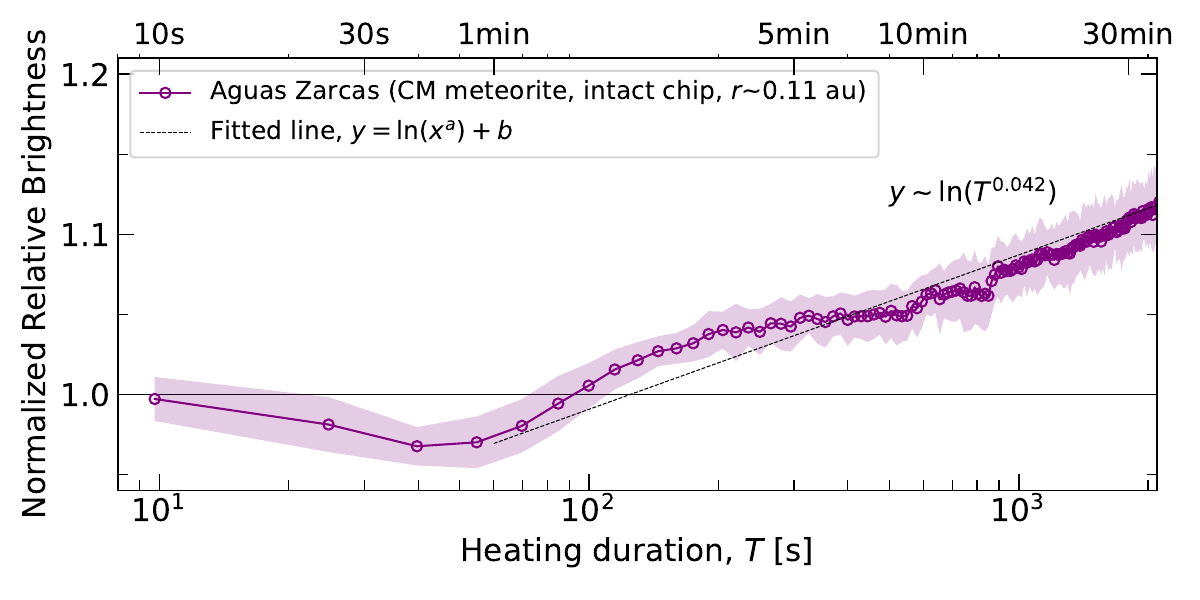}
    \caption{Normalized relative brightness evolution over the heating duration for the CM2 carbonaceous meteorite Aguas Zarcas \citep{Garvie+2021}. The fitted model is shown as a dashed line.}
    \label{fig:meteo-brightness}
\end{figure}

Finally, we suggest that the phenomena observed in our experiments are relevant to real asteroids. We conducted radiation-heating experiments on the CM2 carbonaceous meteorite Aguas Zarcas \citep{Garvie+2021}, following the same procedure as the one used for the simulant. The meteorite showed similar darkening and brightening behavior, although the degree of brightness change differs from CM simulants, likely due to differences in sample preparation. We prepared the CM simulants as powders and sieved them to <45~{\textmu}m, whereas the Aguas Zarcas sample was heated as an intact chip (millimeter scale) without prior grinding (Fig.~\ref{fig:meteo-brightness}).

\subsection{Mechanism of morphology modification and consolidation}
\label{sec:dis-consol}

\begin{figure*}[htbp]
    \centering
    \rotatebox[origin=c]{90}{\textsf{Unheated}}\hspace{0.5em}%
    \raisebox{-0.5\height}{\includegraphics[width=0.85\linewidth]{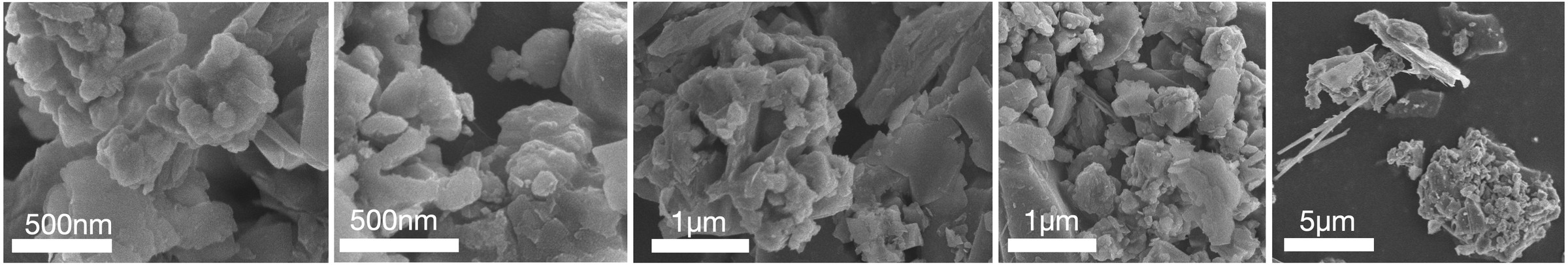}}
    \vspace{1em}

    \rotatebox[origin=c]{90}{\textsf{Heated}}\hspace{0.5em}%
    \raisebox{-0.5\height}{\includegraphics[width=0.85\linewidth]{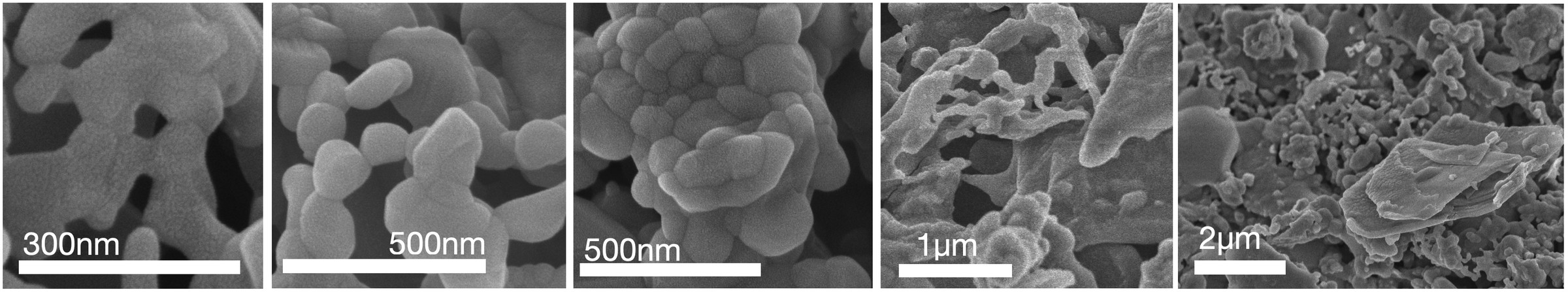}}
    \caption{SEM images of the CM simulant sample (sieved to $<45$~{\textmu}m) before (upper panels) and after (lower panels) heating at $r = 0.11\,$au. Note that the images of the samples before and after heating are not taken from the same location, but are shown at comparable spatial scales. Structures resembling those formed by sintering are observed in the heated sample, whereas such features are absent in the unheated sample. Scale bars are shown below with their corresponding lengths.}

    \label{fig:Neck}
\end{figure*}
Thermal treatment has been suggested to influence the physical properties of asteroid surfaces in both laboratory and observational studies \citep{Delbo+2014, Ito+2018, Molaro+2020, Schirner+2024}. Laboratory experiments have mainly focused on thermal fracturing, a process that generates small particles through repeated thermal cycling caused by diurnal temperature variations on asteroid surfaces \citep{Delbo+2014, Libourel+2020, Patzek+2024, Latsia+2026}. During these cycles, repeated expansion and contraction produce thermal stresses that eventually fracture rocks into smaller grains. On the other hand, observational studies of the NSA Phaethon suggest the presence of relatively coarse surface particles, and sintering has been proposed as a potential mechanism responsible for the formation of such coarse particles \citep{Ito+2018}. Sintering is a thermally activated process in which particles bond together through diffusion or other atomic-scale mass transport mechanisms, forming neck-like bridges between adjacent grains below the melting point of the materials \citep[e.g.,][]{German2010, Chen+2019, Tho+2020}. However, unlike thermal fatigue, no experimental studies have yet confirmed whether radiation heating can induce sintering on carbonaceous NSA surfaces.

In this study, we observe changes in the heated samples that are commonly associated with sintering. First, the heated samples consolidate, which indicates an increase in cohesion. Second, the simulant and meteorite samples shrink after heating, which can indicate a decrease in porosity, although we cannot exclude mass loss because we did not measure the sample mass before and after heating. We note that both changes are assessed qualitatively and not quantified. In addition, SEM images show that the particles in the heated samples are frequently connected by neck-like structures, and that their surfaces appear smoother (Fig. \ref{fig:Neck}). We find no such features in the SEM images we obtained from the fresh samples.  

Additional experimental results are also consistent with the sintering hypothesis. First, we conducted experiments using particles of different sizes: coarse particles ($\sim1000$~{\textmu}m sieve), medium particles ($\sim710$~{\textmu}m sieve), and fine particles (45~{\textmu}m sieve; the primary sample used in this study). Under the same irradiation conditions (i.e., the same $r$), visibly detectable consolidation was observed only in the samples containing fine particles, and those with a larger fraction of fine particles exhibited more pronounced consolidation based on visual inspection (Sect. \ref{sec:consolidation}). This trend is consistent with the tendency for sintering to occur more efficiently in fine-grained materials. Second, by visual inspection, we observed a greater degree of consolidation as $r$ becomes smaller (i.e., stronger radiation flux corresponding to higher heating temperature). Such a tendency is also consistent with sintering being more efficient at higher temperatures. Considering these lines of evidence, together with the shrinkage and the neck-like structures described above, we interpret sintering as a likely mechanism responsible for the morphological modification and consolidation observed in our samples.

Mineral melting is unlikely to be responsible for the smoothing and consolidation observed in our samples. The expected temperatures corresponding to the solar irradiance at $0.09<r<0.19\,$au are not sufficiently high to melt the minerals present in our samples, such as magnetite, olivine, and pyroxene. This interpretation is also supported by the EDS result, which indicates that the granular particles show no significant change in relative elemental composition after heating. Although thermal decomposition or dehydration may occur in some minerals, including serpentine, pyrite, coal, and siderite, these processes do not produce a liquid phase under the expected temperature conditions. Therefore, morphological modification and consolidation caused by melting and resolidification of minerals are unlikely in our experiments. Another possible mechanism for consolidation is mechanical interlocking between grains. Gas species such as H$_2$O and CO released during heating may mobilize particles and potentially promote interlocking between grains. However, microscopic observations do not show clear evidence of interlocking between mineral grains. We therefore consider grain interlocking to be an unlikely explanation for the consolidation observed in our samples.

\begin{figure}[htbp]
    \centering
    \begin{tikzpicture}
        \node[anchor=south west, inner sep=0] (image) at (0,0)
        {\includegraphics[width=0.6\linewidth]{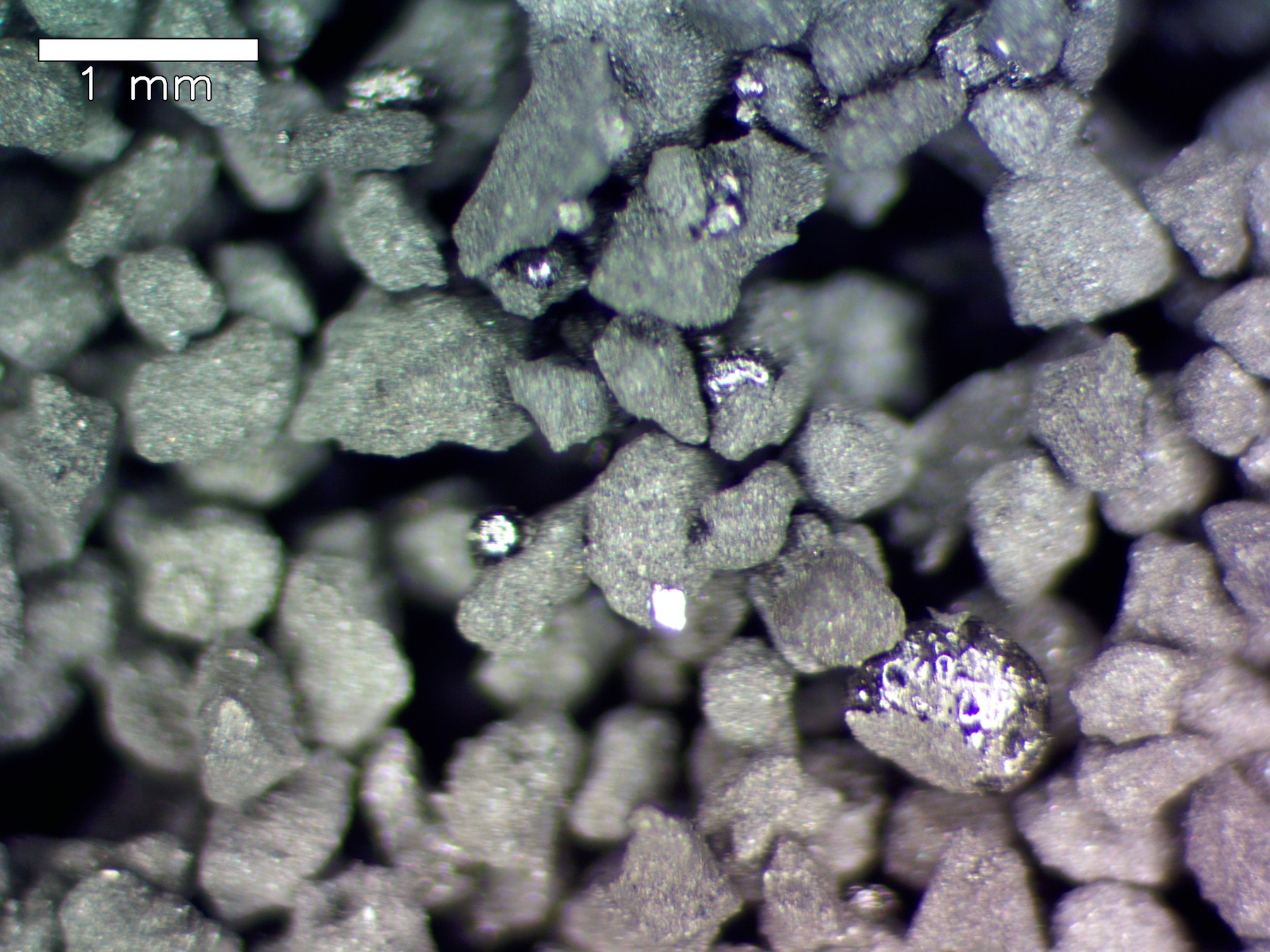}};
        \begin{scope}[x={(image.south east)}, y={(image.north west)}]
            \draw[red, thick, ->] (0.5,0.5) -- (0.56,0.58);
            \draw[red, thick, ->] (0.65,0.14) -- (0.71,0.24);
            \draw[red, thick, ->] (0.31,0.32) -- (0.37,0.4);
            \draw[red, thick, ->] (0.31,0.32) -- (0.37,0.4);
            \draw[red, thick, ->] (0.5,0.67) -- (0.56,0.75);
            \draw[red, thick, ->] (0.45,0.26) -- (0.51,0.34);
        \end{scope}
    \end{tikzpicture}
    \caption{CM simulant with a large grain size after heating. Metallic layer features are observed on the particle surface, as indicated by the arrows.}
    \label{fig:metal-ptl}
\end{figure}

We note an additional observation from experiments using coarse particles ($\sim1000$~{\textmu}m sieve size). After heating, some particles were partially or entirely coated with a thin layer that appeared to have metallic optical properties (Fig.~\ref{fig:metal-ptl}). These shiny particles were not observed in the samples prior to heating. When such particles were crushed manually, the interiors were composed of the original simulant material rather than metallic material. The crushed fragments included flat, plate-like pieces that appeared to originate from the metallic-looking layer, indicating that this material formed a thin coating on the particle surfaces. In this study, we were unable to determine the exact composition of this material. However, it is possible that chemical reactions during heating produced a surface film with metallic optical properties, or that partial melting of a metal-bearing mineral formed a thin coating on the particle surfaces. Therefore, although melting is unlikely to be the primary mechanism responsible for the observed consolidation, we do not completely exclude the possibility that limited melting or chemical alteration may occur during heating.

\subsection{Comparison with observations}

Our experiments indicate that intense radiative heating can modify CM-like materials, but they do not predict the physical state of any given asteroid surface. This is because our experiments do not perfectly reproduce the asteroid's environment. For example, our experiments reproduce heating over laboratory timescales, whereas asteroid surfaces are exposed to repeated diurnal cycling over geological timescales. With this limit in mind, we compare our laboratory results with observational trends among NEAs.

Phaethon is known to have a heterogeneous surface, with reported variations in grain size, roughness, and albedo \citep{MacLennan+2022b, Betzler+2024}. \citet{Ito+2018} attribute the coarse grains to sintering, a hypothesis that has not been constrained experimentally. Our experiments show that consolidation occurs under radiative heating relevant to NSA surfaces. If this
trend applies to Phaethon, its heterogeneity may arise from nonuniform solar heating near perihelion \citep[e.g.,][]{Ohtsuka+2009}, with more strongly
heated regions having undergone greater consolidation and brightening.

\begin{figure}
    \centering
    \includegraphics[width=\linewidth]{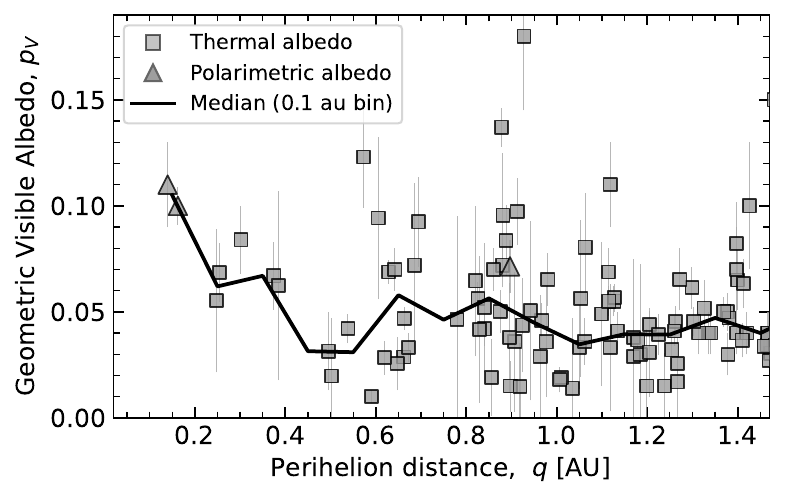}
    \caption{Geometric visible albedo of C-complex asteroids as a function of the perihelion
distance. Albedos are thermal (MP3C archive) or polarimetric \citep{Geem+2022, Ishiguro+2022}. Only objects with $q < 1.6$~au and $\sigma(p_V) < 0.05$ are used for analysis.
The solid line is the median in 0.1~au bins. Objects with $p_V > 0.2$ lie above
the plotted $y$-axis range but are included in the median.}
    \label{fig:albedo_vs_distance}
\end{figure}

Another observational trend can be compared with our results. Published albedos of C-complex asteroids rise inward of $q\sim$0.2--0.4~au (Fig.~\ref{fig:albedo_vs_distance}), driven by the intermediate albedos of Phaethon and 2005~UD ($q=$0.14 and 0.16~au, respectively), the only carbonaceous NSAs whose albedos have been reported. At these distances, such objects are expected to reach peak surface temperatures of about 400--600$^\circ$C \citep{Marchi+2009}, close to the temperature at which our samples and CM meteorites begin to brighten (Fig.~\ref{fig:temp-brightness}). We consider two possibilities for this trend. First, it may be an observational bias that would disappear as more albedos are measured. Second, it may be real, and the scarcity of such objects may itself be physical. \citet{Granvik+2016} and \citet{Tsirvouils+2026} propose that carbonaceous asteroids on small-$q$ orbits are catastrophically disrupted by intense solar heating, which accounts for the small number of known NEAs with $q\lesssim0.2$~au \citep{Toliou+2021}. In either case, the intermediate albedos of two objects that have repeatedly experienced perihelion heating are consistent with our experimental results.

\section{Conclusions}
\label{sec:conclusions}

This work aims to fill a gap in laboratory studies of carbonaceous NSAs, in which the effects of intense radiative heating on surface properties have not been investigated under solar-like irradiation and vacuum conditions. We conducted radiative-heating experiments on CM-type simulants in vacuum to examine changes in surface properties that influence light-scattering behavior, including reflectance and surface texture.

We found that irradiated samples exhibit brightness changes, consolidation, and morphological modifications across spatial scales from sub-micrometer to millimeter. Although the samples are consolidated, they remain mechanically weak and can be easily broken, indicating low cohesive strength comparable to consolidated mud.

We interpret the observed brightness changes as being mainly driven by modifications in the absorption properties of the samples, rather than by changes in surface roughness at the sub-micrometer to micrometer scale. In addition, we consider sintering a likely mechanism for the observed consolidation and morphological modifications. Neck-like structures were identified in SEM images of the heated samples but not in those of the fresh samples. The heated samples became cohesive enough to be lifted with a wire, and they shrank. Consolidation was also stronger at higher temperatures and in finer-grained samples, which is the trend expected for sintering. 

While this study provides a qualitative assessment of radiative heating effects on carbonaceous NSAs, several aspects remain to be explored. For example, our experiments focused on fine-grained samples ($<45$~{\textmu}m), and the effect of grain size was constrained only by the supplementary experiment (Sect. \ref{sec:consolidation}). A more systematic survey across grain sizes is left for future work. In addition, this study considers peak heating conditions, whereas time-dependent thermal evolution, such as diurnal temperature cycling, should be investigated in future work. Finally, quantitative measurements of the mechanical strength of consolidated materials as a function of $r$ would provide further constraints.

These results provide experimental support to help constrain the surface evolution of NSAs in general, and support the sintering hypothesis proposed for (3200) Phaethon in particular \citep{Ito+2018}. This work also provides experimentally derived constraints for interpreting telescopic observations, including reflectance spectra and polarimetric data of carbonaceous NSAs.

\begin{acknowledgements} 
We are grateful to the anonymous referee for the constructive and insightful comments, which have helped improve the quality of this manuscript. JG acknowledges support from the Carl Trygger Foundation (CTS 24:3838). MG acknowledges travel support from Waldemar von Frenckell's foundation. The contributions of L.S., M.G., and A.H. were partially funded by the Swedish Research Council (grant \#2022-04615). AH acknowledges support from the Kempe foundation (JCSMK24-0024). The SEM and EDS studies were carried out using the facilities available at the Atomic Layer Deposition center Finland. This work is based on data provided by the Minor Planet Physical Properties Catalogue (MP3C) of the Observatoire de la Côte d'Azur.

\end{acknowledgements}

\bibliographystyle{aa} 
\bibliography{ref} 

\end{document}